\documentclass[12pt, preprint]{aastex}

\slugcomment{published in \emph{Icarus}, \textbf{207}: 359--372 (2010) | doi: 10.1016/j.icarus.2009.10.013}

\shorttitle{The Evolving Flow of Jupiter's White Ovals and Cyclones}
\shortauthors{Choi et al.}

\begin{document}

\title{The Evolving Flow of Jupiter's White Ovals and Adjacent Cyclones}

\author{David S. Choi and Adam P. Showman}
\affil{Department of Planetary Sciences, The University of Arizona,
				Tucson, AZ 85721}
\email{dchoi@lpl.arizona.edu}

\and

\author{Ashwin R. Vasavada}
\affil{Jet Propulsion Laboratory, California Institute of Technology,
				Pasadena, CA 91109 (U.S.A.)}

\begin{abstract}
We present results regarding the dynamical meteorology of Jupiter's White Ovals at different points in their evolution. Starting from the era with three White Ovals FA, BC, and DE (\emph{Galileo}), continuing to the post-merger epoch with only one Oval BA (\emph{Cassini}), and finally to Oval BA's current reddened state (\emph{New Horizons}), we demonstrate that the dynamics of their flow have similarly evolved along with their appearance. In the \emph{Galileo} epoch, Oval DE had an elliptical shape with peak zonal wind speeds of $\sim$90 m s$^{-1}$ in both its northern and southern peripheries. During the post-merger epoch, Oval BA's shape was more triangular and less elliptical than Oval DE; in addition to widening in the north-south direction, its northern periphery was 20 m s$^{-1}$ slower, and its southern periphery was 20 m s$^{-1}$ faster than Oval DE's flow during the \emph{Galileo} era. Finally, in the \emph{New Horizons} era, the reddened Oval BA had evolved back to a classical elliptical form. The northern periphery of Oval BA increased in speed by 20 m s$^{-1}$ from \emph{Cassini} to \emph{New Horizons}, ending up at a speed nearly identical to that of the northern periphery of Oval DE during \emph{Galileo}. However, the peak speeds along the southern rim of the newly formed Oval BA were consistently faster than the corresponding speeds in Oval DE, and they increased still further between \emph{Cassini} and \emph{New Horizons}, ending up at $\sim$140 to 150 m s$^{-1}$. Relative vorticity maps of Oval BA reveal a cyclonic ring surrounding its outer periphery, similar to the ring present around the Great Red Spot. The cyclonic ring around Oval BA in 2007 appears to be moderately stronger than observed in 1997 and 2001, suggesting that this may be associated with the coloration of the vortex. The modest strengthening of the winds in Oval BA, the appearance of red aerosols, and the appearance of a turbulent, cyclonic feature to Oval BA's northwest create a strong resemblance with the Great Red Spot from both a dynamical and morphological perspective.

In addition to the White Ovals, we also measure the winds within two compact cyclonic regions, one in the \emph{Galileo} data set and one in the \emph{Cassini} data set. In the images, these cyclonic features appear turbulent and filamentary, but our wind field reveals that the flow manifests as a coherent high-speed collar surrounding relatively quiescent interiors. Our relative vorticity maps show that the vorticity likewise concentrates in a collar near the outermost periphery, unlike the White Ovals which have peak relative vorticity magnitudes near the center of the vortex. The cyclones contain several localized bright regions consistent with the characteristics of thunderstorms identified in other studies. Although less studied than their anticyclonic cousins, these cyclones may offer crucial insights into the planet's cloud-level energetics and dynamical meteorology.
\end{abstract}


\section{Introduction}

Oval BA, a large, anticyclonic vortex in Jupiter's southern hemisphere, is the latest stage in the evolution of the former White Ovals, a group of three atmospheric vortices that were approximately centered at 30$^{\circ}$S latitude (planetocentric). Their origin dates back to the late 1930's, when an entire latitudinal band unexpectedly clouded over and became bright white \citep{Peek58}, eventually coalescing into three discrete, anticyclonic vortices. Further observations of the vortices revealed that their longitudinal width decreased over time, in addition to repeatedly approaching and receding away from each other \citep{Rogers95}. \citet{Youssef03} have speculated that the apparent stability and longevity of the White Ovals was a consequence of an alternating configuration of cyclones and anticyclones known as a classical von K\'arm\'an vortex street. Such a configuration would inhibit vortex mergers because the cyclonic systems that developed between each White Oval would prevent the anticyclones from combining. \citet{Vasavada98} appraised the dynamics of one of the original three White Ovals (Oval DE) using \emph{Galileo} image pairs and determined that the vortex was a compact oval with maximum winds near 100 m s$^{-1}$. \citet{Simon98} extended the analysis of the \emph{Galileo} images by measuring the velocities of White Oval BC, and determined its top winds to be $\sim$120 m s$^{-1}$. Both of these velocity assessments are referring to measurements of individual wind vectors, and their overall results are similar to measurements made using data from the \emph{Voyager} epoch \citep{Mitchell81, Dowling88, Legarreta05}, implying that the vortex was stable over decadal timescales. However, \citet{Simon98} also noted that Oval BC became rounder between \emph{Voyager} and \emph{Galileo} as its latitudinal extent increased and its longitudinal extent decreased.

Although a strong cyclonic feature was present between Ovals BC and DE in 1997, it apparently dissipated by 1998, allowing the two ovals to merge. The merger did not destroy the two vortices but instead yielded a new vortex (Oval BE) that was larger than either BC or DE but less than their areas combined \citep{Sanchez-Lavega99}. Though previous observational studies reported that larger vortices occasionally absorb smaller vortices, perhaps providing the energy needed by large, long-lived vortices on Jupiter for maintenance against dissipation (\citet{Vasavada05}; \citet{Li04}), a merger between vortices both of similar size and of the length scale of a White Oval had not been previously observed. After the initial merger, observers spotted a new cyclonic structure (labeled O1) near the new Oval BE. \citet{Sanchez-Lavega99} surmised that O1 was either a pre-existing structure that had survived the merger, or that it was a product of the merger, as its area plus the area of the new Oval BE was similar to the areas of the two former White Ovals. Eventually, as the other remaining White Oval (FA) approached Oval BE, the cyclonic structure O1 was either pushed aside or slowly sheared apart, and the remaining pair of White Ovals merged to form Oval BA \citep{Sanchez-Lavega01}.

In late 2005, amateur observers first reported that Oval BA had reddened, leading many to dub the vortex as the ``Little Red Spot.'' \citet{Garcia-Melendo09} chronicles a full description of Oval BA's long-term morphological evolution and interactions with the Great Red Spot using both spacecraft and amateur images. Soon after the color change, \citet{Simon-Miller06} examined Hubble Space Telescope (HST) imagery of Oval BA and found evidence for an acceleration of the vortex's flow. \citet{Cheng08a} analyzed \emph{New Horizons} images of Oval BA during the spacecraft's flyby of Jupiter and reported additional evidence for a substantial strengthening of the flow in Oval BA, with top wind measurements at $\sim$180 m s$^{-1}$. However, \citet{Marcus07} have suggested that the dynamics did not change sufficiently to account for the color change, and further suggested that an overall climate change was responsible for the coloration. Recent analysis of \emph{Cassini} and HST images of Oval BA taken before and after its color change by \citet{Asay-Davis09} seem to support the idea that the vortex's dynamics remained relatively unchanged, as their measurements indicate that the velocities within the flow in Oval BA did not change appreciably after the coloration event. Another analysis of the \emph{New Horizons} data by \citet{Hueso09} also did not find evidence of a strengthening flow when compared against earlier images of the vortex from \emph{Cassini} and HST. Thus, there is disagreement regarding the latest measurements of Oval BA's winds and what role they played in the coloration of the vortex, which is especially unfortunate given their timing at a critical point in the vortex's evolution.

In this paper, we report results from our assessment of Oval BA's winds starting from its previous state as a White Oval up to its current condition as a compact, red anticyclone. Our effort is a comparative study of multi-mission data regarding the vortex's dynamical meteorology using a common semi-automated technique for measuring its winds. Such study is necessary for robust inferences on vortex evolution in order to constrain what role dynamics had in Oval BA's changing appearance.

\section{Methodology}

\subsection{Data Sets}

We obtained all of the data used in this chapter from various NASA Planetary Data System (PDS) nodes. Table \ref{Table: ovalba_data} summarizes our data sets. Unless explicitly stated otherwise, all latitudes quoted in this paper are planetocentric, and all longitudes are positive west.

\subsubsection{\emph{Galileo} E6 orbit}

We acquired navigated and calibrated mosaics of Oval DE, BC, and an intervening cyclone\footnote{These mosaics are available online at\\ \texttt{http://atmos.nmsu.edu/PDS/review/Jupiter/Galileo\_Maps/E/E6}}. During its E6 orbit in February 1997, \emph{Galileo} observed these vortices using its Solid State Imager \citep{Belton92}. Mission planners designed this observation sequence to examine the cloud structure and dynamics of the White Ovals using multiple camera filters. We only used images that were taken with the 756 nm near-infrared continuum filter. This filter is useful for imaging opacity variations (cloud contrast features) located within an ammonia cloud deck \citep{Banfield98}. \emph{Galileo} observed the vortices four times during the E6 orbit; the first and second mosaics in our data set are separated by 38m 25s, and the second and third mosaics are separated by 85m 57s. We did not include in our analysis a fourth mosaic, composed using images taken one planetary rotation (about 10 hours) prior to the set of three mosaics, because the cloud contrast features in Jupiter's atmosphere experience sufficient rotation, shear, and inherent variability during the relatively long image separation time that our image tracker cannot reliably recognize these features and measure winds. 

The raw \emph{Galileo} images were calibrated and despiked with the VICAR software package developed at the Jet Propulsion Laboratory. One of the authors (ARV) determined the navigational pointing data for the images. The mosaics were composed using MaRC, an open-source map-making software package\footnote{MaRC is available for download at \texttt{http://sourceforge.net/projects/marc} for interested readers.}. The mosaics are a cylindrical projection of the images that cover the range 22--47$^{\circ}$S and 135--60$^{\circ}$W. All images were taken near the nadir of the spacecraft. The mosaics are mapped at a resolution of 0.018$^{\circ}$ pixel$^{-1}$ ($\sim$20 km pixel$^{-1}$), which slightly oversamples the original data, which is at a resolution of $\sim$23 km pixel$^{-1}$. Further details about these mosaics and the processing steps can be found in \citet{Vasavada98} and online at the PDS Atmospheres node.

\subsubsection{\emph{Cassini} flyby}

\emph{Cassini} observed the newly formed Oval BA on 1 January 2001 using its Imaging Science Subsystem \citep{Porco04} soon after its closest approach to Jupiter. At this point in the system's evolution, the mergers had already completed, but Oval BA had not changed color. \emph{Cassini} obtained two images of Oval BA nearly 53 minutes apart with the CB2 continuum band filter, which has a central transmission wavelength of 750 nm. We used the CISSCAL software package \citep{Porco04} to calibrate the raw images, and other custom software to navigate and map the images. The observation was relatively wide-angle compared to the other images that we analyzed, as a single frame captured the Oval and its surroundings, as well as a planetary limb for absolute navigation; thus, no mosaicking was necessary. We mapped the images using a rectangular projection at a resolution of 0.05$^{\circ}$ pixel$^{-1}$ ($\sim$60 km pixel$^{-1}$, essentially reproducing the original resolution of the images. 

\subsubsection{\emph{New Horizons} flyby}

Near its closest approach with Jupiter, \emph{New Horizons} observed a reddened Oval BA on 27 February 2007 using its Long-Range Reconnaissance Imager (LORRI) \citep{Cheng08b}. LORRI does not employ filters but observes over a broad spectrum of visible wavelength light (350--850 nm) in order to optimize its signal when \emph{New Horizons} arrives at Pluto. This broad band observation captures somewhat more light from diffuse hazes located at a higher altitude than the main cloud deck. However, this added contribution is not expected to adversely affect our wind measurements, which typically follow the motion of the distinct cloud contrast features within the main ammonia cloud deck. \citet{Li06} found insignificant differences in the overall wind pattern results when examining \emph{Cassini} ISS images that utilize various filters in the visible portion of the spectrum.  

LORRI observed Oval BA using a square 2x2 imaging sequence to capture each quadrant of the vortex and its surroundings. The two sequences are separated by 30 minutes. We obtained the calibrated LORRI image frames from the Planetary Data System. The calibration pipeline consisted of bias subtraction, flatfield correction, and absolute calibration. The spacecraft's close range to Jupiter during the observations enable us to map the vortex at a relatively high resolution ($\sim$15 km pixel$^{-1}$) with a rectangular projection. Unfortunately, the observation sequence did not capture a planetary limb for absolute navigation to a latitude/longitude grid. When mosaicking the images, we assume that the lower-left frame of each mosaic has the correct navigation using the provided spacecraft pointing data. The other frames in the mosaic are then navigated relative to the ``correct'' frame by correlating overlapping regions captured in the frames. Because the center of the vortex will have stationary winds, we eliminate the residual relative navigational error between mosaics by aligning the mosaics so that the center of the vortex is stationary between frames. Any inherent drift in the Oval will also be removed by aligning the mosaics in this manner, but Oval BA's drift rate ($\sim$2.5 m s$^{-1}$, as recently measured by \citet{Garcia-Melendo09}) does not have a significant effect on the navigation of the vortex given the relatively short time interval (30 minutes) between imaging sequences. We determine the center of the Oval by a visual inspection of the circulation through a two-frame animation, or blinking, of the image pair. We are confident that we have sufficiently removed the navigation errors, as the location of the calm circulation center found in our subsequent velocity map matches the location of the circulation center that we defined visually.

\begin{table}
\begin{center}
\begin{tabular}{lclr}
\hline
Data Set & Resolution & Time Interval &Spacecraft Clock Time \\
 & Original (Projected) & & \\
 & [km pixel$^{-1}$] & & \\
\hline \hline

\emph{Galileo} E6 & 23 (20) & 38m 25s  & 383548622--383619422 \\
 & & 85m 57s &  \\
 & & 124m 22s &  \\
\emph{Cassini} flyby & 60 (60) & 52m 46s & 1357028060--1357031226 \\
\emph{New Horizons} flyby & 15 (15) & 30m  & 34851719--34853729 \\

\hline
\end{tabular}

\vspace{0.75 cm}

\begin{tabular}{lcc}
\hline
Data Set & High-Pass Filter Kernel & Correlation Box \\
 & & Large (Small)\\
\hline \hline

\emph{Galileo} E6 &0.522$^{\circ}$ & 1.08$^{\circ}$ (0.18$^{\circ}$) \\
\emph{Cassini} flyby & 0.55$^{\circ}$ & 2$^{\circ}$ (1$^{\circ}$) \\
\emph{New Horizons} flyby & 0.504$^{\circ}$ & 0.8$^{\circ}$ (0.08$^{\circ}$) \\

\hline
\end{tabular}

\vspace{0.5 cm}

\caption{\label{Table: ovalba_data}Details regarding the image data and analysis techniques used in this study. The spacecraft clock time column lists the range of times for the images comprising the data set according to the spacecraft's internal clock, each of which has an independently arbitrary zero point. The resolutions listed refer to the original and map-projected resolutions of the images. The time intervals listed for the \emph{Galileo} data set refer to the separation time between mosaics 1 and 2, 2 and 3, and 1 and 3, from top to bottom. The sizes of the high-pass filter kernel and the feature tracking correlation boxes that we used during data processing correspond for each side. The kernel and correlation boxes are square.}
\end{center}
\end{table}

\subsection{Analysis Techniques} 

In all of the component images, we removed the variance in the observed brightness seen in the mosaics by dividing the brightness by the cosine of $\mu_o$, the local solar incidence angle at each pixel. This normalized the images and enhanced areas that were located near the planet's terminator. Furthermore, because absolute photometry is unimportant for our dynamical study, we then processed each image through a high-pass filter in order to enhance the cloud contrast features. The high-pass filter employs a square boxcar kernel when enhancing the images; the sizes of the square kernels used for processing each mosaic are listed in Table \ref{Table: ovalba_data} and were kept as consistent as possible. 

We apply our semi-automated cloud feature tracker \citep{Choi07} to measure winds. We assume that the imaged cloud contrast features are passive tracers of the winds. We also assume that the measured motions are confined to the NH$_3$ cloud deck and that no motions from significantly different layers of the atmosphere are captured in the images. Our software extracts a portion of the earlier image in an image pair and defines it as a basis correlation window. This window is then compared against portions of the later image in the image pair that are in the vicinity of the basis window. Our algorithm determines the comparison of correlation windows with the highest statistical cross-correlation score, and computes the spatial displacement between the two portions. This measurement is then used to calculate the wind velocity, and the resulting wind vector is placed at the center of the original basis correlation window. Once each measurement is complete, this process repeats systematically throughout the entire image, with the location of the basis window shifting by a discrete amount after each iteration. The amount of spacing between wind vector measurements is $\sim$0.05$^{\circ}$ for the \emph{Galileo} and \emph{Cassini} data, and $\sim$0.03$^{\circ}$ for the \emph{New Horizons} data.

Our feature tracking software utilizes two sizes of square correlation windows in determining the wind vector. Initially, the software uses a large correlation box to measure the synoptic motion. The results from this first measurement are set as an initial estimate for the wind vector. The software then repeats the measurement with a small correlation box to further refine the measurement. The wind measurement from the small correlation box is kept unless this measurement deviates strongly from the initial estimate, in which case the initial measurement is kept. We test various sizes for the large and small boxes across all three imaging data sets and subjectively choose box sizes that optimize the quality of the subsequent results by minimizing the amount of uncertainty and frequency of spurious wind vectors. These sizes are listed in Table \ref{Table: ovalba_data}. The spurious wind vectors that can be generated by our software are easily recognizable upon visual inspection. In a grayscale pixel map representing wind magnitude, the pixels denoting unreliable vectors are extreme light or dark values juxtaposed against neighboring pixels. Typically these results occur along image edges or in areas with degraded image quality (such as areas with visible mosaic seams). We manually removed those results before proceeding with our analysis, and these spurious results do not affect our overall conclusions.

There are two principal sources of uncertainty in the magnitude and direction of our calculated wind vectors. Imprecise information about the spacecraft's position and orientation along with uncertainties in the camera pointing knowledge produce uncertainty in the wind measurements from navigational uncertainty. We have manually corrected the spacecraft pointing and reduced this error contribution to be at or below 1 pixel, and the propagated uncertainty onto the wind measurements is not expected to be significant. Thus, the main contributor to the uncertainty in our results should be attributable to the locational uncertainty in the velocity vectors. Our software assigns the location of each measured wind vector to the center of the correlation box, regardless of the box size, but the feature(s) determining the final vector could lie anywhere within the box. Assuming horizontal wind shear exists, this induces an uncertainty in the actual wind speed at the center of the box. The amount of uncertainty is primarily controlled by the sizes of the correlation boxes used in the feature tracking algorithm. We estimate that the maximum uncertainty for an individual wind measurement is 10--20 m s$^{-1}$ using the software parameters in this current work (typical values should be $<$10 m s$^{-1}$), and overall, we do not expect that the uncertainties significantly affect our conclusions. We also note that for the vast majority of results, the cloud contrast features are discrete objects that occupy a significant fraction of each correlation window. The scenario that would result in the most uncertainty (an isolated feature located at one corner of the window in a high shear environment) is expected to be uncommon.

Unlike the \emph{Cassini} and \emph{New Horizons} data sets, which comprise only two mosaics each and thus only one possible feature tracking comparison within each set, the \emph{Galileo} data set comprises three mosaics, which yields three possible image pairs for feature tracking. We performed all three comparisons and averaged the results together to create the final wind vector field. However, before averaging, we removed spurious results (typically caused by image seams) from the three individual comparisons; thus, not all locations in the final, averaged wind vector field for the \emph{Galileo} data set have contributions from all three feature tracking comparisons. We filled in any locations with missing data in the final wind vector fields for each of the three epochs in this study using nearest-neighbor averaging (defining the value of the wind at the location of a missing pixel to be the average of the values in its 8 neighboring pixels).

\section{Results}

\subsection{Wind Vector Maps}

We performed a visual inspection of the vortices in order to define a central meridian and latitude for the Ovals based on their cloud patterns. Our examination also defined the edges of the vortices from the visible cloud morphology in order to gauge vortex sizes. We calculate vortex eccentricities by using the lengths of the semi-major and semi-minor axes defined by our visual assessment. Table \ref{Table: oval_quants} summarizes the values for the central points for the vortices as well as their lengths.

\begin{figure}[htb]
  \centering
  \includegraphics[width=5.5in, keepaspectratio=true]{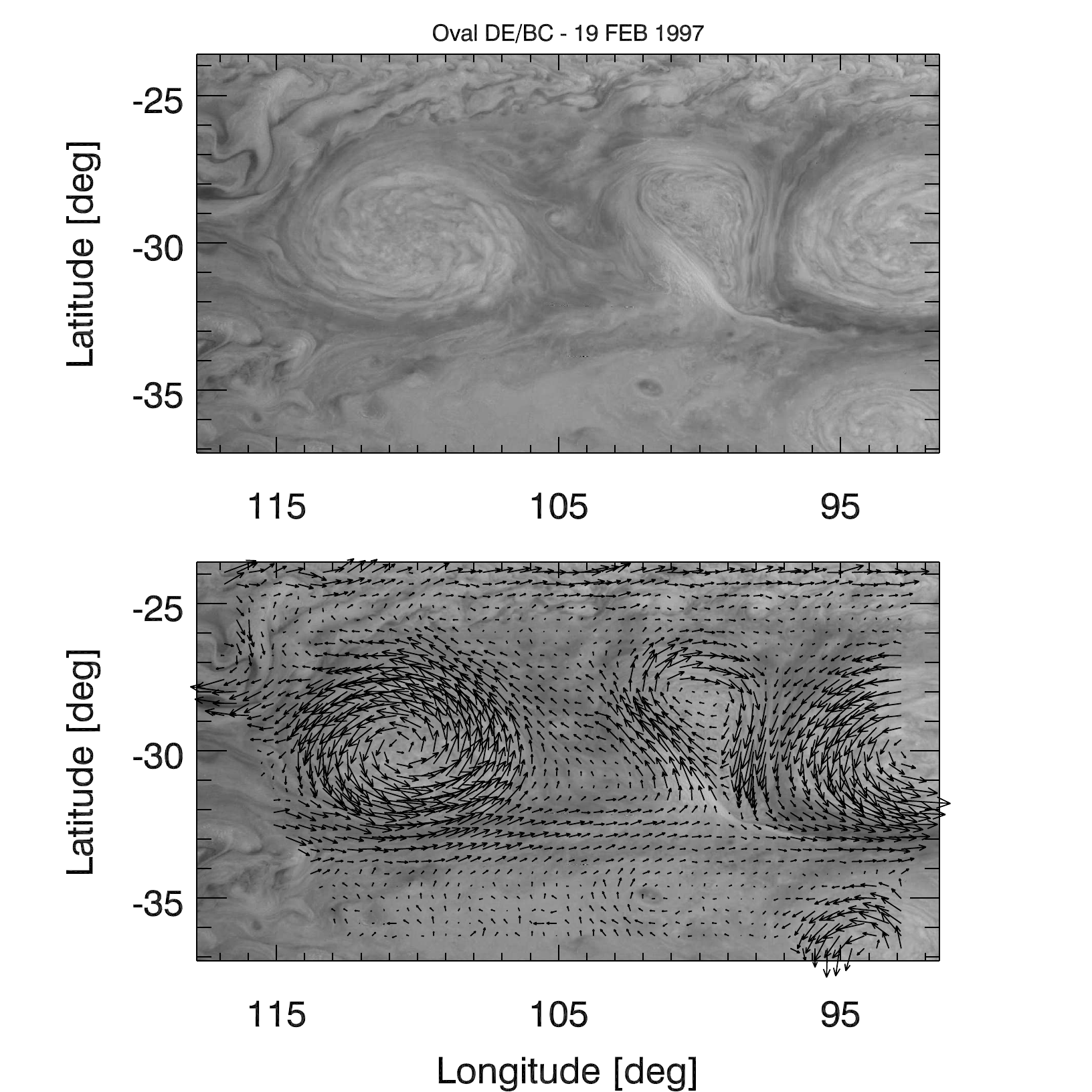}
  \caption[Oval DE and surrounding region with overlain wind vectors, February 1997]{
    \label{Figure: e6_windvec}
    (top) Oval DE (left), cyclone (middle), and a portion of Oval BC (right) as imaged by \emph{Galileo} in February 1997. Image contrast has been enhanced for clarity. (bottom) Image with overlain wind vectors. Vectors have been smoothed using nearest-neighbor averaging of vectors within 0.25$^{\circ}$ latitude or longitude of the selected grid point. Less than 2\% of the total wind vectors in our data set are shown for clarity.
    }
\end{figure}

Figure \ref{Figure: e6_windvec} shows our measured wind vectors corresponding to Oval DE and BC along with a teardrop-shaped cyclone overlain on one of the \emph{Galileo} mosaics. We examine Oval DE's cloud structure and measure its spatial extent to be approximately 8.5$^{\circ}$ x 6$^{\circ}$ (9,000 x 7,000 km), with an aspect ratio of 1.24. (We define aspect ratio as the ratio of the major (east-west) axis length with the minor (north-south) axis length.) We estimate its central latitude to be 29.5$^{\circ}$S. Many of the details regarding the jets noted by \citet{Vasavada98} match what is seen in our current results. For example, the image is bounded at the north by the eastward jet centered at 24$^{\circ}$S, though our results only show the southern edge of this jet. Furthermore, the eastward jet between 32 and 33$^{\circ}$S is also visible, with its northern half appearing to loop around the intervening cyclone between DE and BC. The notable exception to the overall general match with \citet{Vasavada98} in the qualitative jet descriptions is our possible detection of the weak westward zonal jet at 29$^{\circ}$S between Ovals DE and BC that was not mentioned in the previous work. Around 104--105$^{\circ}$W longitude, we detect a possible portion of this westward jet as it is begins to deflect north of DE. However, the exact identity of this flow is unclear. It could simply be a component of the westward jet, although this is uncertain since the 29$^{\circ}$S westward jet to the east of Oval BC is not visible in our mosaics. An alternative possibility is that the flow in question is not part of a jet but is instead part of the northwestward flow at the cyclone's southwest quadrant that branches off and eventually becomes incorporated into DE's flow. The westward zonal jet is more visible as it flows away from DE at its western edge near 29.5$^{\circ}$S, 115$^{\circ}$W.

\begin{figure}[htb]
  \centering
  \includegraphics[width=6.5in, keepaspectratio=true]{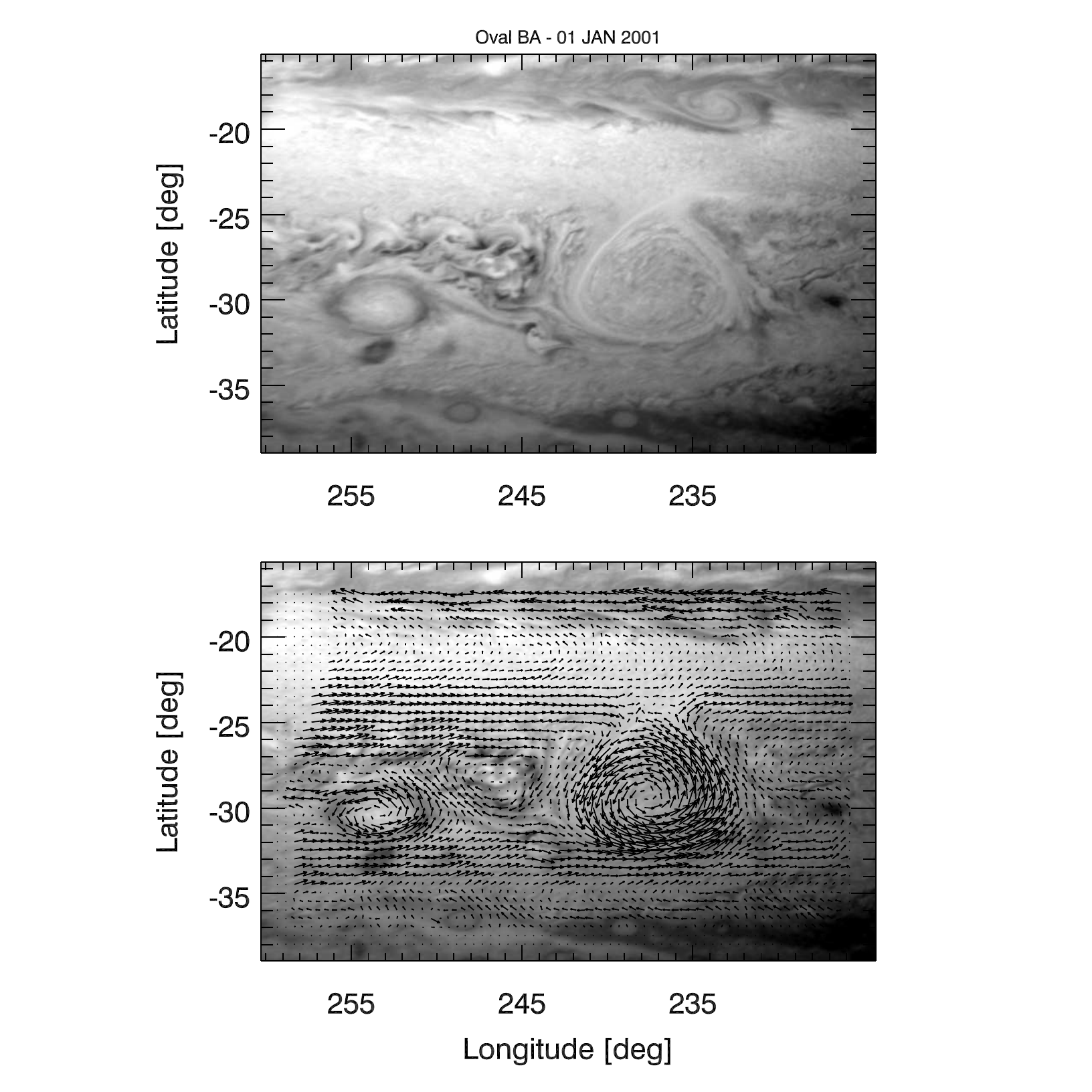}
  \caption[Oval BA and surrounding region with overlain wind vectors, January 2001]{
    \label{Figure: iss_windvec}
    (top) Oval BA as imaged by \emph{Cassini} in January 2001. Image contrast has been enhanced for clarity. (bottom) Image with overlain wind vectors. Vectors have been smoothed using nearest-neighbor averaging of vectors within 0.25$^{\circ}$ latitude or longitude of the selected grid point. Less than 2\% of the total wind vectors in our data set are shown for clarity.
    }
\end{figure}

Figure \ref{Figure: iss_windvec} is a wind vector map from the \emph{Cassini} flyby data of the newly formed Oval BA along with a compact anticyclone to its west. The maximum axes for Oval BA measure nearly 9$^{\circ}$ x 8$^{\circ}$, with a reduced aspect ratio equal to 1.04. Its central latitude, as estimated by eye, is 29.25$^{\circ}$S, almost unchanged from DE's position nearly four years prior. The overall shape of Oval BA appears remarkably different compared with its appearance during the \emph{New Horizons} epoch and with Oval DE or BC's appearance during the \emph{Galileo} epoch. The outer edge of the vortex is no longer ellipsoidal but instead approximates a teardrop or triangle. The unusual shape may be a transient effect of the White Oval mergers, which had just completed, or it could be an effect originating from the interaction from the vortex and the 24$^{\circ}$S eastward zonal jet stream bounding it to its north. \citet{Garcia-Melendo09} have performed numerical simulations of Oval BA with the initial latitude of the vortex as one free parameter. Their results suggest that the latitude of Oval BA affects its dynamical shape, and have found that when Oval BA is at the latitude in which it resided during the \emph{Cassini} flyby, its shape becomes more triangular, and when it drifts to the south, like where it ultimately wound up during the \emph{New Horizons} flyby, its shape is more symmetric and ellipsoidal.

The relatively wider-angle perspective provided by this image yields insight into how the background zonal jets interact with the vortices. At the northern boundary of this image is the strong westward jet located at 18$^{\circ}$S. The two prominent eastward jets at 24$^{\circ}$S and 33$^{\circ}$S are also obvious in Figure \ref{Figure: iss_windvec}. However, the triangular shape of Oval BA may be affecting the flow of the jet at 24$^{\circ}$S, as its northern tip appears to cause some of the jet's flow to be diverted southward around the vortex (239$^{\circ}$W, 25$^{\circ}$S). The re-connection of the flow with the eastern half of the jet (235$^{\circ}$W, 24$^{\circ}$S) appears to be associated with a bright white cloud feature that appears to branch off from the vortex and become entrained in the jet stream. It is possible to trace out the westward jet at 29$^{\circ}$S flowing southwestward from the northwestern quadrant of Oval BA south of a cyclonic feature west of BA and then to the north of a compact anticyclone. The jet is also seen at the eastern edge of the image and appears to deflect completely to the north and merge with the eastward jet at 24$^{\circ}$S. The jet deflection appears similar to ``circulating currents'' noted during observations of disturbances in Jupiter's atmosphere that temporarily disrupt the alternating jet stream pattern \citep{Rogers95}. However, unlike the recirculating currents that developed east of the Great Red Spot as part of a South Tropical Zone Disturbance (one example is shown in detail in \citet{Smith79}, with measurements of this recirculating current discussed in \citet{Legarreta05}), it does not appear likely that a similar atmospheric disturbance is causing the deflection east of Oval BA in this case. Instead, the recirculating pattern is likely the result of the simple interaction of the jet with the vortex. Another possibility is that we are not observing the jet at all, but instead a loosely organized, weak cyclonic vortex that is accompanying Oval BA to its east as part of a von K\'arm\'an vortex street. This would certainly fit the pattern of vortices already present in the region.

\begin{figure}[htb]
  \centering
  \includegraphics[width=7in, keepaspectratio=true]{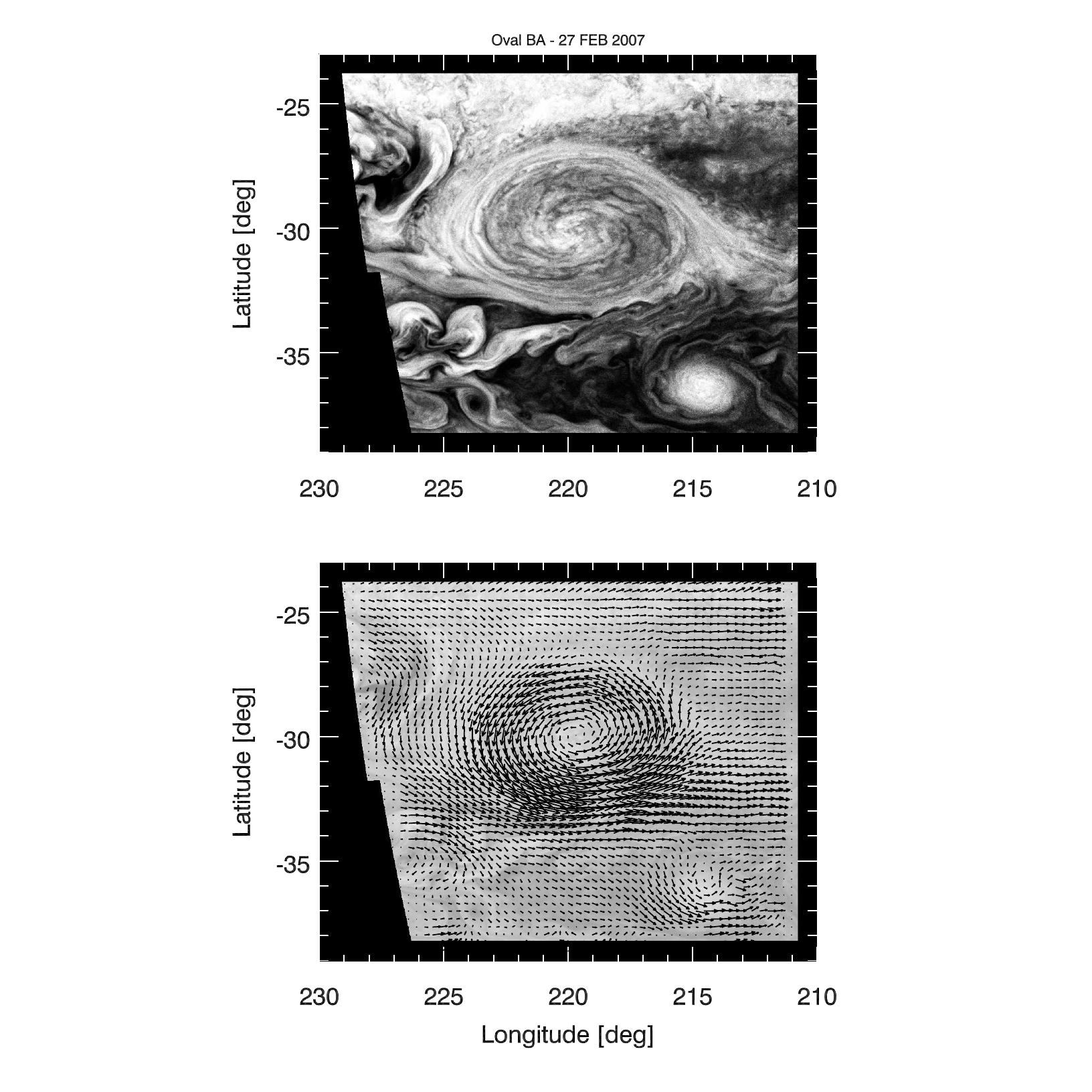}
  \caption[Oval BA and surrounding region with overlain wind vectors, February 2007]{
    \label{Figure: lorri_windvec}
    (top) Oval BA as imaged by \emph{New Horizons} in February 2007. Image contrast has been enhanced for clarity. (bottom) Image with overlain wind vectors. Vectors have been smoothed using nearest-neighbor averaging of vectors within 0.25$^{\circ}$ latitude or longitude of the selected grid point. Less than 2\% of the total wind vectors in our data set are shown for clarity.
    }
\end{figure}

Our final wind vector map, from \emph{New Horizons}, is shown as Figure \ref{Figure: lorri_windvec}. In its most recent epoch, Oval BA has evolved back into its classic ellipsoidal shape. The zonal extent of the clouds associated with the vortex appears to have expanded considerably: we estimate its size at 11$^{\circ}$ x 7$^{\circ}$. Its aspect ratio also significantly increased to 1.49. The central latitude of the system during this observation is estimated to be at almost 30$^{\circ}$S (with the caveat that we are unable to absolutely navigate the images because a limb is not present in the data set), indicating that the Oval has drifted slightly southward. This southward drift matches the overall conclusion regarding the latitudinal position of the vortex made by \citet{Garcia-Melendo09} from a more complete set of Oval BA observations. Both of the jets at 24$^{\circ}$S and 33$^{\circ}$S are clearly evident. However, in contrast with the \emph{Cassini} observations, the jet at 24$^{\circ}$S appears largely unaffected by BA, whereas the jet at 33$^{\circ}$S is slightly deflected to the south. Unlike our \emph{Galileo} and \emph{Cassini} observations, we do not detect the weak westward jet at 29$^{\circ}$S. 

\subsection{Wind Profiles}

\begin{figure}[htb]
  \centering
  \includegraphics[width=6in, keepaspectratio=true]{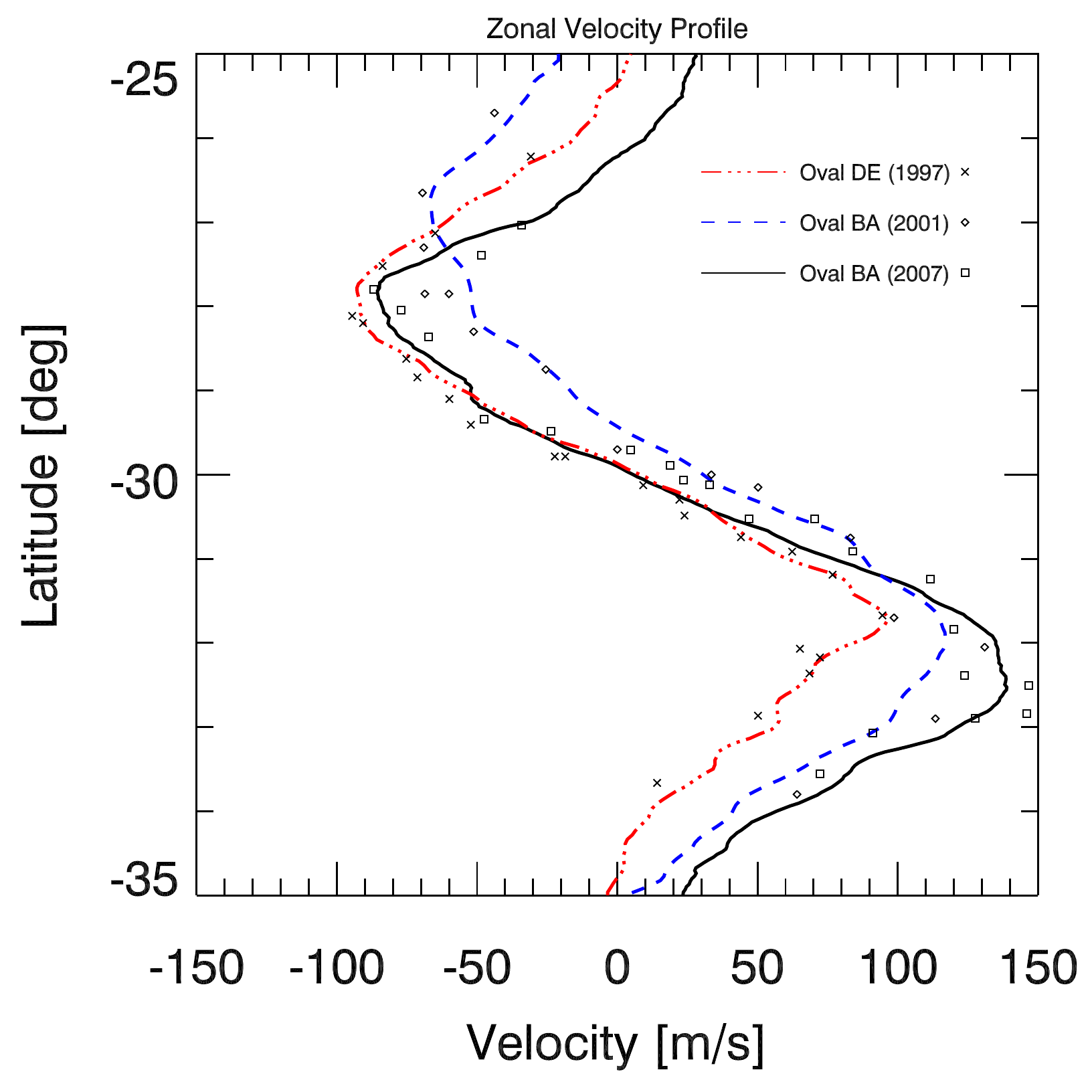}
  \caption[Zonal Wind Profile for Ovals DE/BA]{
    \label{Figure: uvel_profile}
    Zonal wind profile for Ovals DE and BA throughout the epoch studied in this paper. The zonal component of the wind is averaged in 0.25$^{\circ}$ latitude bins over 2$^{\circ}$ longitude centered at the Oval's central meridian (shown in Table \ref{Table: oval_quants}). Manual wind measurements for all three epochs are also shown with corresponding symbols. 
    }
\end{figure}

\begin{figure}[htb]
  \centering
  \includegraphics[width=6in, keepaspectratio=true]{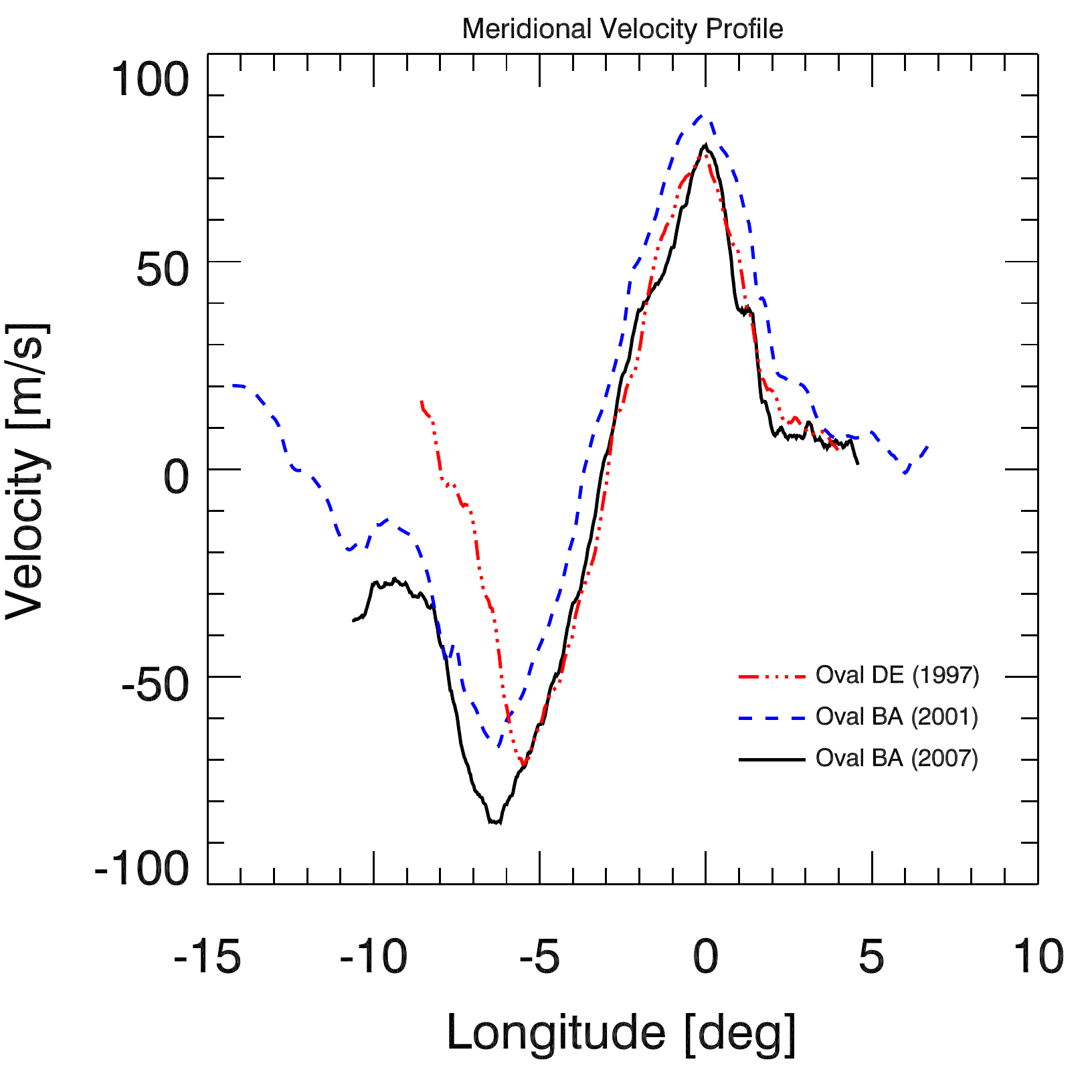}
  \caption[Meridional Wind Profile for Ovals DE/BA]{
    \label{Figure: vvel_profile}
   Meridional wind profile for Ovals DE and BA throughout the epoch studied in this paper. The meridional component of the wind is averaged in 0.25$^{\circ}$ longitude bins over 2$^{\circ}$ latitude, centered at the Oval's central latitude (shown in Table \ref{Table: oval_quants}). The zero longitude for each profile is defined as the location of the peak northward velocity, and longitude is positive east for this figure only.
    }
\end{figure}

Figures \ref{Figure: uvel_profile} and \ref{Figure: vvel_profile} display zonal and meridional wind profiles for Ovals DE and BA. We constructed these profiles by first using the central meridians and latitudes for the Ovals shown in Table \ref{Table: oval_quants}. We then took wind measurements within 1$^{\circ}$ of the central axis and averaged them in 0.25$^{\circ}$ latitude (longitude) bins for the zonal (meridional) wind profile. We only use the zonal and meridional components of the wind vectors when creating these profiles. We have also supplemented the zonal wind profile in Figure \ref{Figure: uvel_profile} with manual measurements to demonstrate the fidelity of our automated method. When assuming a 1 pixel error in our manual measurements, this corresponds to $\sim$5 m s$^{-1}$ uncertainty for each measurement in the Oval DE (1997) and Oval BA (2007) data sets, and a $\sim$20 m s$^{-1}$ uncertainty for each measurement in the Oval BA (2001) data set. Other sources of error (navigational, etc.) would contribute additional uncertainty to the measurements.

Our zonal wind profile provides evidence for a substantial increase in speed between \emph{Galileo} and \emph{New Horizons} within the southern portion of the Ovals' flow collar, even when we consider any residual uncertainty in the navigation of the \emph{New Horizons} data. We also note from Figure \ref{Figure: uvel_profile}, that the peak values for the zonal wind from the wind profile in the southern half of Oval BA in 2007 is now comparable to the peak values from the profile of the southern half of the Great Red Spot in 2000 \citep{Choi07}. In contrast to the southern half of the vortex, the northern portion of the flow collar does not change as dramatically: the measurements of Oval BA from 2007 show that the flow is quite similar to the flow seen in Oval DE in 1997 in the northern half and does not show stronger flow like in the south. However, the decrease in velocity at the northern half of the collar between 1997 and 2001 is likely associated with Oval BA's shape, as the growth of the vortex to its north can be clearly seen in the \emph{Cassini} wind profile. A possible consequence of the vortex's new shape is the interaction of the 24$^{\circ}$S jet with the flow of Oval BA, producing a weaker current in the flow collar than previously measured in 1997. The subsequent increase in the speed of the northern flow collar between 2001 and 2007 could be an effect of the vortex reverting back to a more stable configuration (note the similarities in the northern half of the 1997 and 2007 profiles). This process of restoring the vortex's shape back to its more ellipsoidal form may have played a role in increasing the velocity of the vortex and/or transporting chromophores, though further studies are necessary to determine if the timing of the coloration event and the evolution in the dynamics is more than a coincidence. 

When comparing our zonal wind profile of Oval DE with the profile reported by \citet{Vasavada98}, we show agreement in the shapes of the profiles and the magnitude of the peak velocity at the northern collar. The peak values along the transects in Figures \ref{Figure: uvel_profile} and \ref{Figure: vvel_profile} are listed in Table \ref{Table: oval_quants}. The peak velocity at the southern collar in the current profile is $\sim$10 m s$^{-1}$ less than the peak value measured by \citet{Vasavada98}. The methods for constructing the profiles are identical, and the central meridians for Oval DE defined in both studies are similar (110.3$^{\circ}$W vs. their 110.5$^{\circ}$W). The difference could be the result of implementing a higher resolution grid for extracting our wind vectors, or because the current profile averages results from three separate feature track comparisons (\citet{Vasavada98} only used one image pair). A similar discrepancy occurs with the \emph{Cassini} zonal wind profile of Oval BA reported in \citet{Simon-Miller06}; they report the peak zonal wind along their profile to be $>$ 150 m s$^{-1}$, which is $\sim$40m s$^{-1}$ above our results. We speculate that the cause of this difference may lie in the sheer quantity of velocity vectors in our data set compared to the number of vectors manually tracked by \citet{Simon-Miller06}, and that our data set thus includes more measurements of slower flow.

We observe strengthening in the western portion of Oval BA when comparing the meridional wind profile (Figure \ref{Figure: vvel_profile}) before and after its color change. However, this increase in velocity for the western portion from 2001 to 2007 may also be associated with the interaction of the 24$^{\circ}$S jet at the northern tip of Oval BA in 2001 and its effects downstream in the flow, or could be associated with a variety of other effects resulting immediately from the final White Oval merger. A slight weakening is seen in the eastern portion, though the peaks in the eastern portion of the profile are all within 10 m s$^{-1}$ of one another. We also note that there is remarkably little change in the horizontal (east-west) length of the vortex when using the metric of the peak-to-peak length in the meridional wind profile. This is in contrast to the significant lengthening observed when assessing the Oval's length based on its visible clouds. The contrasts in length between the cloud morphology and the velocity profile suggest some discretion when interpreting visible changes in vortex clouds as a marker for changes in the physical flow \citep{Shetty07}. Continued caution is warranted when relying solely on these metrics when assessing a vortex's dynamical state as a vortex's influence only ends when its flow matches that of the ambient flow.

Nevertheless, we note several evolving characteristics of Oval DE/BA when examining both its cloud morphology and its velocity profiles. The remarkable trait about Oval DE's velocity profile is that it is quasi-symmetrical: the zonal velocities along its northern and southern sides are similar, and the meridional velocities along the eastern and western sides are similar, but $\sim$20 m s$^{-1}$ slower than the zonal flows, as shown in Table \ref{Table: oval_quants}. This symmetry is broken in Oval BA, as the southern portion of the flow is typically stronger than the rest of the vortex. There is a $\sim$50 m s$^{-1}$ difference in the zonal wind magnitudes between the northern and southern portions of Oval BA that is consistent in both 2001 and 2007. However, the inherent magnitudes of the velocity profile peaks themselves increase by nearly 20 m s$^{-1}$ from 2001 to 2007. This increase in the velocities may be associated with Oval BA's color change. When examining both its cloud aspect ratio and velocity profiles, we note a decrease in the Ovals' aspect ratios between \emph{Galileo} and \emph{Cassini}, followed by an increase between \emph{Cassini} and \emph{New Horizons}. We summarize all of these quantitative characteristics in Table \ref{Table: oval_quants}.

\begin{table}
\begin{center}
\begin{tabular}{|l|c|c|c|c|c|}

\hline
\multicolumn{6}{|c|}{\textbf{Cloud Morphology}} \\
\hline
 & Cent. Lat & Cent. Lon & E-W Length & N-S Length & Aspect Ratio \\
\hline \hline

Oval DE (1997) & 29.5$^{\circ}$S & 110.3$^{\circ}$W & 8.5$^{\circ}$ (9.1 $\times$ 10$^3$ km) & 6.2$^{\circ}$ (7.4 $\times$ 10$^3$ km) & 1.24 \\
\hline
Oval BA (2001) & 29.3$^{\circ}$S & 237.3$^{\circ}$W & 8.7$^{\circ}$ (9.3 $\times$ 10$^3$ km) & 7.8$^{\circ}$ (8.9 $\times$ 10$^3$ km) & 1.04 \\
\hline
Oval BA (2007) & 30.0$^{\circ}$S & 219.6$^{\circ}$W & 11.0$^{\circ}$ (11.7 $\times$ 10$^3$ km) & 6.8$^{\circ}$ (7.8 $\times$ 10$^3$ km) & 1.49 \\
\hline

\end{tabular}

\vspace{0.75 cm}

\begin{tabular}{|l|c|c|c|}

\hline
\multicolumn{4}{|c|}{\textbf{Wind Profiles}} \\
\hline
 & E-W Length & N-S Length & Aspect Ratio \\
\hline \hline

Oval DE (1997) & 5.5$^{\circ}$ (5.9 $\times$ 10$^3$ km) & 3.8$^{\circ}$ (4.4 $\times$ 10$^3$ km) & 1.32 \\
\hline
Oval BA (2001) & 6.3$^{\circ}$ (6.7 $\times$ 10$^3$ km) & 5.3$^{\circ}$ (6.1 $\times$ 10$^3$ km) & 1.11 \\
\hline
Oval BA (2007) & 6.3$^{\circ}$ (6.7 $\times$ 10$^3$ km) & 4.7$^{\circ}$ (5.5 $\times$ 10$^3$ km) & 1.23 \\
\hline

\end{tabular}

\vspace{0.75 cm}

\begin{tabular}{|l|c|c|c|c|c|}

\hline
\multicolumn{6}{|c|}{\textbf{Peak Wind Profile Velocities (m s$^{-1}$)}} \\
\hline
 & $u$ (N) & $u$ (S) & $v$ (W) & $v$ (E) & Peak Indiv. Meas. \\
\hline \hline

Oval DE (1997) & -92.9 & 96.3 & -71.9 & 76.7 & 104.1 \\
\hline
Oval BA (2001) & -66.8 & 117.0 & -66.9 & 85.4 & 122.5 \\
\hline
Oval BA (2007) & -85.5 & 138.9 & -85.2 & 78.0 & 150.6 \\
\hline

\end{tabular}

\vspace{0.5 cm}

\caption{\label{Table: oval_quants}Lengths and eccentricities of Oval DE and Oval BA using criteria based on their wind profiles or on a visual inspection of their cloud morphology. We define aspect ratio as the ratio of the major (east-west) axis length with the minor (north-south) axis length. Aspect ratios may appear to be slightly off due to rounding of the listed measurements. The listed wind profile velocities are the values of the peak zonal (meridional) velocities along the central latitudinal (longitudinal) profile of the vortex. The peak individual wind vector measurements reported are the highest wind velocity values measured in the smoothed velocity maps.}
\end{center}
\end{table}

\subsection{Velocity Magnitudes}

\begin{figure}[htb]
 \centering
  \includegraphics[width=4in, keepaspectratio=true]{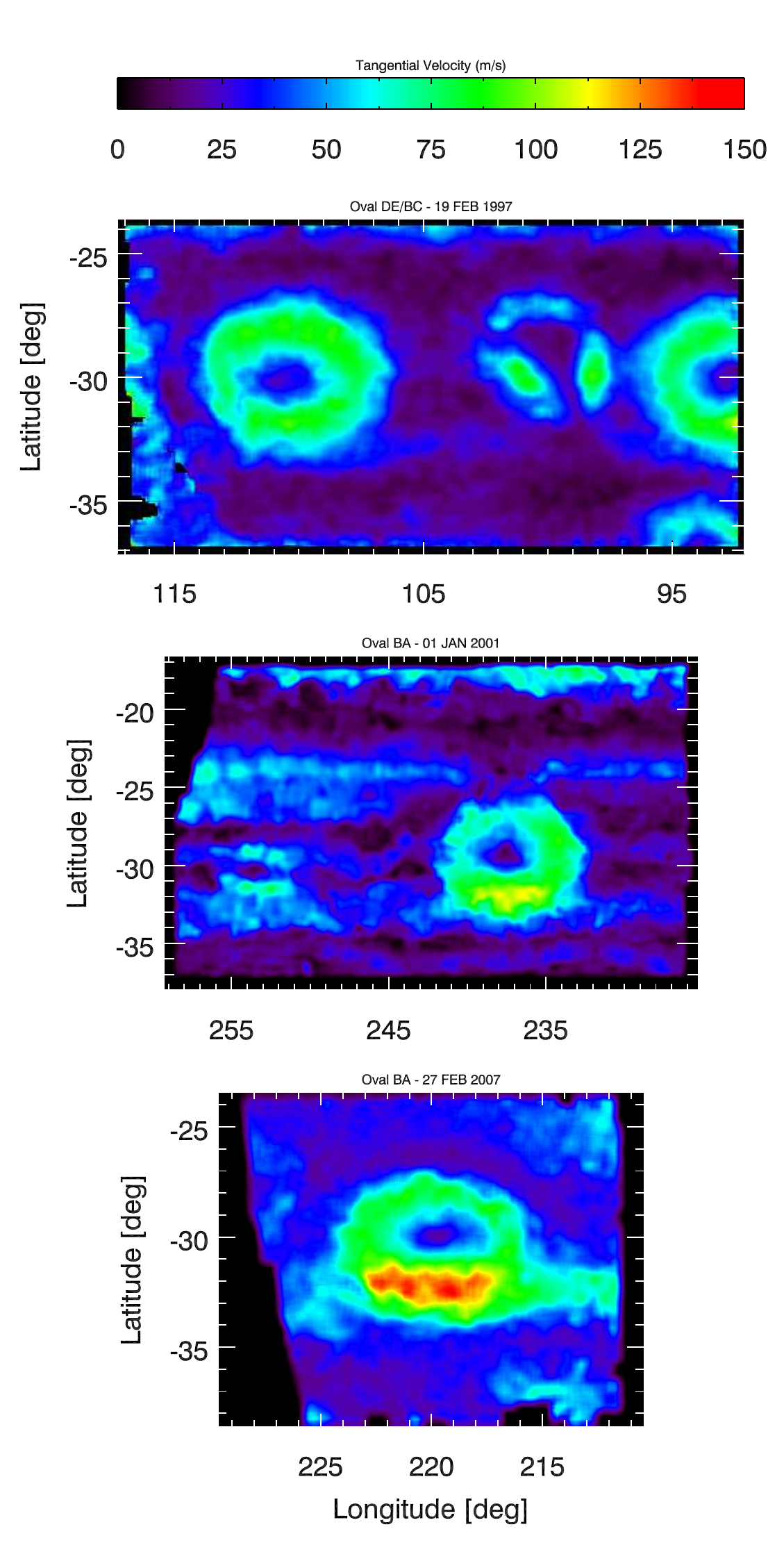}
  \caption[Velocity Magnitude Map for Ovals DE/BA]{
    \label{Figure: velmap}
    Map of velocity magnitude for Ovals DE/BA from 1997--2007. The velocity map has been smoothed using nearest-neighbor averaging of data within 0.25$^{\circ}$ latitude or longitude of each pixel.
    }
\end{figure}

Figure \ref{Figure: velmap} is a map of velocity magnitude $|\mathrm{v}|$, where $|\mathrm{v}|= \sqrt{u^2 + v^2}$, and $u$ and $v$ are the zonal and meridional velocities, respectively. We smoothed the map using nearest-neighbor averaging of data within 0.25$^{\circ}$ latitude or longitude of each pixel. Though it may appear from Figure \ref{Figure: velmap} that the interiors of the Ovals are relatively calm, our velocity profiles (Figures \ref{Figure: uvel_profile} and \ref{Figure: vvel_profile}) reveal that velocity simply increases with radial distance (i.e. angular velocity is essentially constant). This structure contrasts with the structure of the Great Red Spot, where the flow is mostly confined into an outer collar \citep{Marcus93}, and where its interior region slowly flows in the \emph{opposite} direction as the outer flow collar \citep{Vasavada98, Choi07}. During the \emph{Galileo} epoch, the structure of Oval DE appears relatively symmetric, with the northern portion of the flow collar slightly stronger than the remainder of the collar. The structure of BA from \emph{Cassini} onward is remarkably different; the southern portion of the flow collar is clearly stronger than the other three quadrants. We also note moderate increases in the flow's velocity before and after Oval BA's color change in the southern portion of the vortex. These increases are modest in magnitude ($\sim$20--30 m s$^{-1}$) but notable for the areal extent in which these increases appear to have taken place within the vortex, as seen when comparing the velocity maps of Oval BA in 2001 (\emph{Cassini}) with velocity maps of Oval BA in 2007 (\emph{New Horizons}) in Figure \ref{Figure: velmap}. This somewhat confirms observations of fast winds reported by \citet{Simon-Miller06} and \citet{Cheng08a}, but not to the magnitude previously reported by these studies. Table \ref{Table: oval_quants} lists the maximum values of individual velocity measurements in the smoothed maps, which are slightly lower than previous studies; \citet{Simon-Miller06} and \citet{Cheng08a} report 180 and 172 m s$^{-1}$, respectively, as maximum tangential wind values for Oval BA from Hubble Space Telescope and \emph{New Horizons} data. In our raw, unsmoothed wind measurements, we report that some isolated areas of BA appear to be flowing near 180--190 m s$^{-1}$ during \emph{New Horizons}. However, these regions occupy only a very small fraction of Oval BA's southern collar, and often manifest as individual pixels in our full-resolution wind vector map, suggesting that these sharp spikes in speed could be a result of random measurement noise. As shown in Figure \ref{Figure: velmap}, typical peak speeds throughout the bulk of the southern collar are $\sim$140--150 m s$^{-1}$. The maximum individual wind measurements in the smoothed velocity maps for Oval DE during the \emph{Galileo} era, and Oval BA during the \emph{Cassini} flyby (104 and 122 m s$^{-1}$, respectively) are considerably lower than the \emph{New Horizons} Oval BA measurements.

\subsection{Cyclones}

Cyclones are an enigmatic component of Jupiter's atmosphere. Jovian anticyclones are typically easier to distinguish, as they organize themselves into coherent, compact oval structures (spots). More than 90\% of the visible spots in Jupiter's atmosphere are anticyclonic \citep{MacLow86, Li04}. Cyclonic spots, in contrast, are relatively rare and are usually noted as ``brown barges'' in tropical latitudes \citep{Hatzes81}. Typically, cyclonic areas on Jupiter are zonally elongated, filamentary structures that may be multi-lobed \citep{Morales-Juberias02}. \citet{Simon98} and \citet{Youssef03} highlight the significant role that cyclones have had in the evolution of the White Ovals. Furthermore, \citet{Legarreta05} performed a quantitative study regarding the vorticity and velocity of cyclones and anticyclones from \emph{Voyager} and \emph{Galileo} imagery and determined that the vorticity of cyclones, like their anticyclonic counterparts, increased towards their periphery. Furthermore, the magnitude of the vorticity tended to be controlled primarily by their latitudinal location and not by other physical characteristics such as the size of the vortex or by the meridional wind shear conditions present surrounding the vortex.

\begin{figure}[htb]
  \centering
  \includegraphics[width=5.5in, keepaspectratio=true]{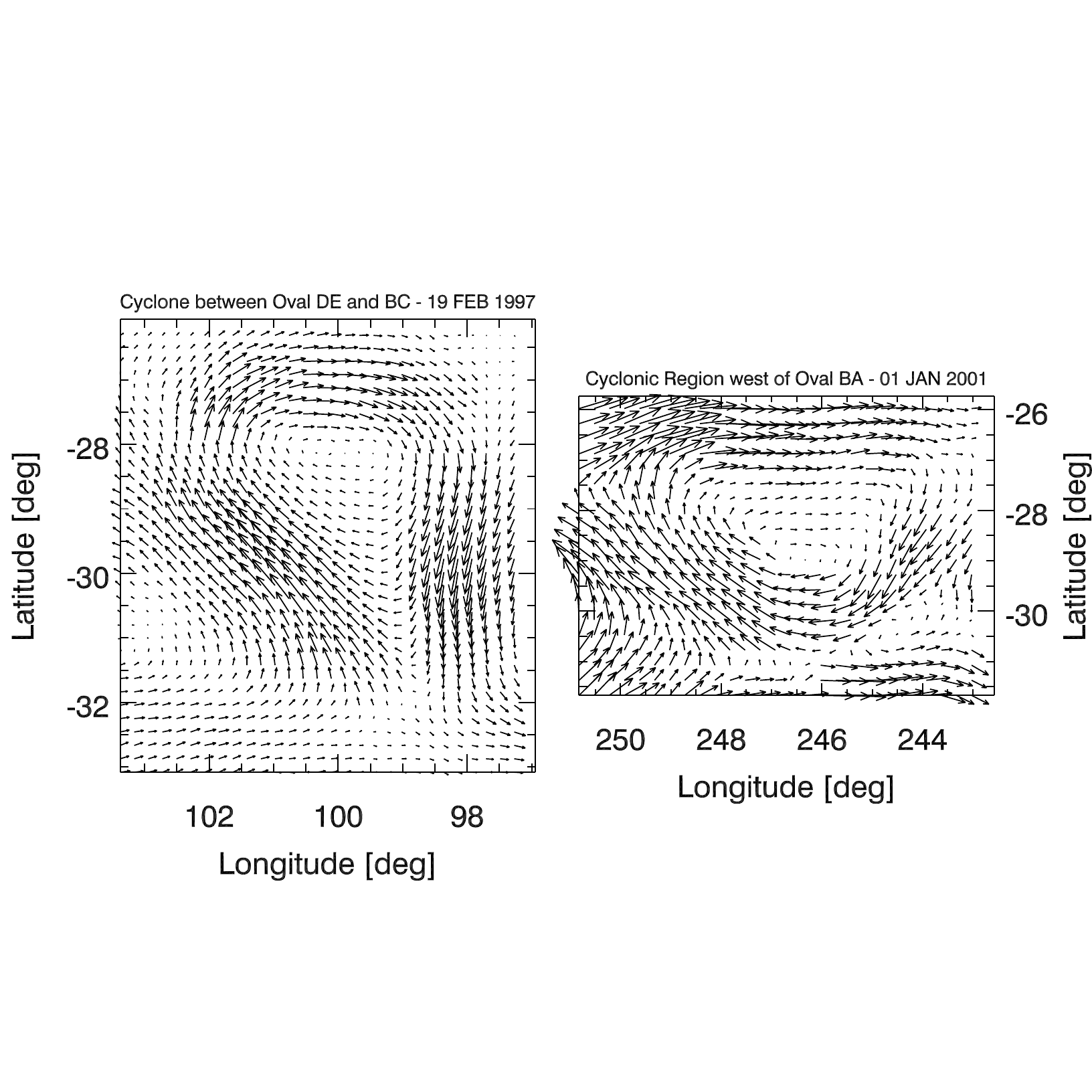}
  \caption[Wind Vector Map of Cyclonic Regions]{
    \label{Figure: cyclone_windvec}
    Wind vector map showing the flow in the cyclone in between Oval DE and BC during \emph{Galileo} (left) and in the cyclone region located west of Oval BA during \emph{Cassini} (right). For this figure, wind vectors have been smoothed using nearest-neighbor averaging of vectors within 0.25$^{\circ}$ latitude or longitude of each grid point. For clarity, we show $\sim$6\% (left) and $\sim$3\% (right) of the wind vectors in our data set.
    }
\end{figure}

We examine the wind flow of the two prominent cyclones imaged in our mosaics (Figure \ref{Figure: cyclone_windvec}). The presence of a particularly large, coherent cyclone imaged by \emph{Galileo} between White Ovals DE and BC is striking, but also critical to prevent the Ovals from merging with each other \citep{Youssef03}. This particular interloping cyclone is large, encompassing an area approximately 5$^{\circ}$ x 4$^{\circ}$. However, its shape is irregular, with a wide, northern elliptical lobe attached to a southern blunted tip. The cyclone's flow structure is also somewhat peculiar. A meridional wind profile taken across the cyclone's central latitude reveals its velocity peaks to be 55.1 and -83.8 m s$^{-1}$ on its western and eastern flanks, respectively. However, the peak meridional wind on the western flank is somewhat misleading, as the flow there is more northwestward instead of directly northward. Velocity magnitudes in that region typically range between 80 and 100 m s$^{-1}$. The northern portion of the cyclone flows somewhat slower, with a peak zonal wind at 54.9 m s$^{-1}$. Overall, the velocity magnitudes of the cyclonic vortex imaged by \emph{Galileo} on its eastern and western sides are comparable to the velocity magnitudes for its two neighboring anticyclones, as shown in Figure \ref{Figure: velmap}. Eventually, this cyclone either dissipated or failed to obstruct the White Ovals, allowing DE and BC to merge shortly after the \emph{Galileo} E6 observations.

In the \emph{Cassini} observations, a cyclonic feature appears to the west of Oval BA and to the east of a small, compact anticyclone. This cyclonic feature appears filamentary and turbulent, but our wind field reveals that the flow is cyclonic and relatively laminar. \citet{Marcus04} demonstrated through his numerical simulations that the physical appearance of the clouds are not necessarily representative of the flow characteristics of the underlying vortex. In his simulations, a compact, oval cyclone would generate filamentary and chaotic cloud patterns. This cyclonic feature appears to fit that description. We also note the similarity in shape with the cyclone observed by \emph{Galileo}: both vortices crudely resemble an upside-down triangle, with a wider extent in its northern half and a blunt tip to its south. One key difference is that the cyclonic feature imaged by \emph{Cassini} extends further to the north and appears to be interacting with the southern portions of the westward jet at 24$^{\circ}$S. 

\begin{figure}[htb]
  \centering
  \includegraphics[width=5.5in, keepaspectratio=true]{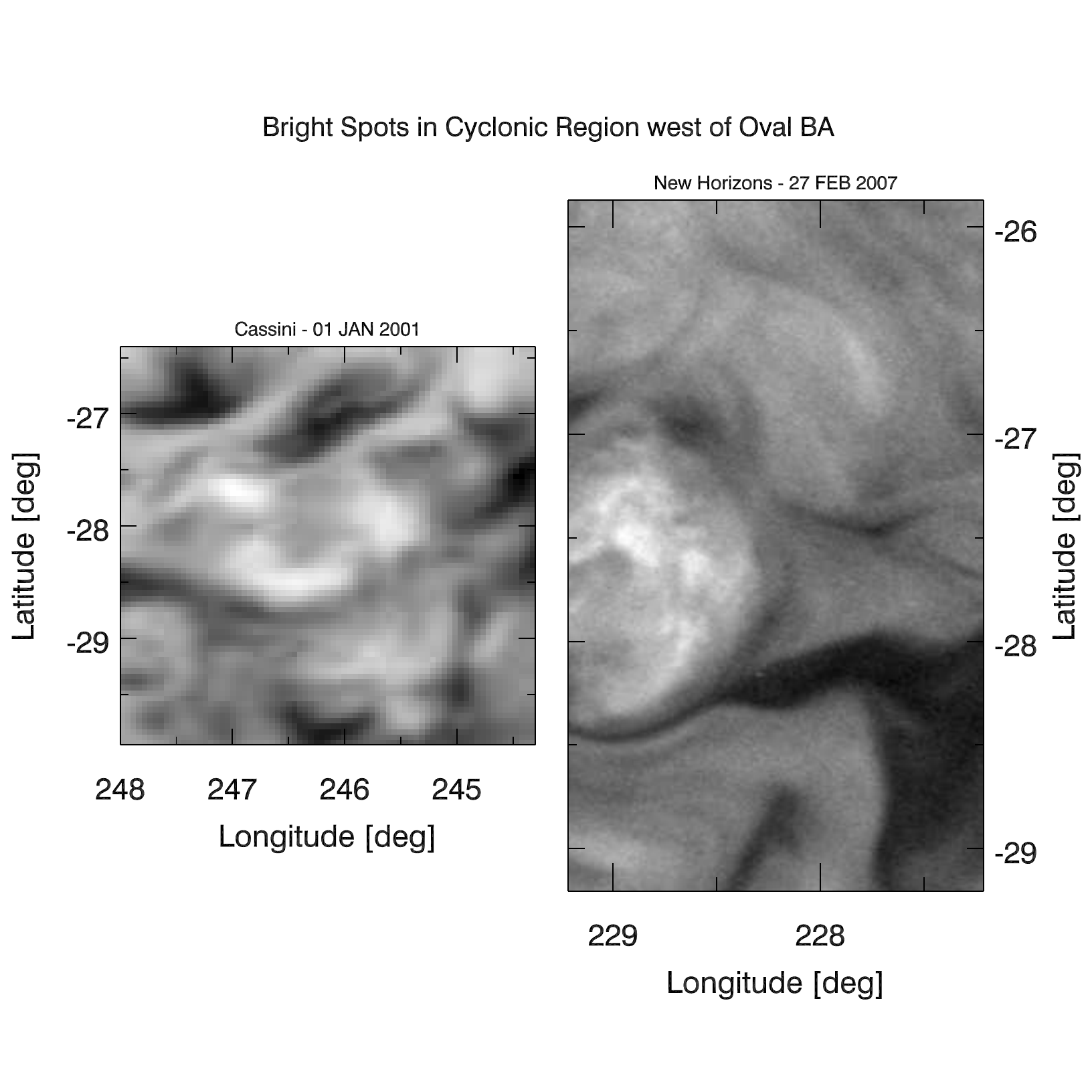}
  \caption[Bright Clouds in Cyclonic Regions west of Oval BA]{
    \label{Figure: cyclone_bright}
    Images of the bright clouds imaged in the cyclonic region west of Oval BA during \emph{Cassini} (left) and \emph{New Horizons} (right). Image contrast has been enhanced in this figure for clarity.
    }
\end{figure}

We also note the presence of bright white clouds in the interior of this cyclonic feature (Figure \ref{Figure: cyclone_bright}). These bright clouds do not appreciably change in morphology or size in the relatively short time interval between images. These clouds are among the brightest pixels in the entire map and could be thunderstorms. \citet{Gierasch00} previously observed thunderstorms and night-side observations of lightning within the storms using \emph{Galileo} images. The lightning identified by \citet{Gierasch00} was located in a cyclonic feature west of the Great Red Spot. \citet{Dyudina04} conducted a search for lightning occurring in \emph{Cassini} flyby images but did not recognize this cyclonic feature west of Oval BA as a location where lightning was spotted in subsequent nightside images. However, we note that their study identified bright, white clouds seemingly embedded within the turbulent wake region northwest of the GRS where associated lightning had been detected in nightside images. The bright, white clouds in the GRS turbulent wake appear to be similar to the clouds at the center of the cyclonic feature west of Oval BA, as they are clearly local maxima in albedo and are comparable in shape and size. 

The \emph{New Horizons} images only partially observe the cyclonic feature northwest of Oval BA. Our wind field confirms that the outer periphery of the region is also circulating cyclonically. Its overall morphology is again similar with what is seen in the \emph{Cassini} images: folded, turbulent ribbons of clouds along with bright, white localized clouds (Figure \ref{Figure: cyclone_bright}). These clouds are again among the brightest pixels in the entire map and presumably could be thunderstorms. Unfortunately, we cannot determine how these clouds evolve in the 30-minute interval between images; these clouds only appear in the later (second) image in the \emph{New Horizons} image pair as the earlier (first) image does not extend sufficiently to the west. A search for lightning during the \emph{New Horizons} flyby of Jupiter \citep{Baines07} also does not identify the region northwest of Oval BA as a location with observed lightning flashes. However, it appears that their study was focused on observing higher latitudes and may have neglected mid-latitudes. We also note an extensive region with filamentary, ribbon-like clouds southwest of Oval BA that also exhibits relatively smooth, coherent flow. The latest observations of Oval BA from a GRS/BA conjunction in July 2008\footnote{\texttt{http://hubblesite.org/gallery/album/entire\_collection/pr2008027a/}} reveal that the cyclonic feature northwest of BA remains present and may be a long-term feature of the entire system. Overall, the apparent longevity of this region, the localized bright spots (possible thunderstorms), and the overall folded, turbulent morphology of the clouds are evidence that the cyclonic feature northwest of Oval BA is very much analogous to the turbulent wake region northwest of the Great Red Spot. 

\subsection{Vorticity}

We implement the approach of \citet{Dowling88} to measure vorticities directly from the wind field maps. Because we only measure horizontal flow, we can only assess the vertical component of vorticity. Relative vorticity $\zeta$ is defined as

\begin{equation} 
\zeta = -\frac{1}{R}\:\frac{\partial u}{\partial \phi} + \frac{u}{r}\:\sin\:\phi + \frac{1}{r}\:\frac{\partial v}{\partial \lambda} 
\end{equation}

where $r$ and $R$ are the zonal and meridional radii of curvature, $\phi$ is planetographic latitude, and $\lambda$ is longitude, positive east. (The longitude in our images and results are positive west and were converted to positive east for the vorticity calculation.) Absolute vorticity $\eta$ is simply $\eta = \zeta + f$, where $f$ is the Coriolis parameter $2 \Omega \sin \phi$. The equation for vorticity requires calculation of the spatial derivatives for zonal and meridional velocities. When constructing the maps for these quantities, we extract a finite quantity of vectors within 0.5$^{\circ}$ in both latitude and longitude around each map pixel. Following \citet{Dowling88}, we fit the extracted vectors to a linear function in $\lambda$ and $\phi$. This smoothing simplifies the calculation of the spatial derivatives, which are determined from the slope of these linear fits. Once the vorticity and divergence quantities are calculated, the algorithm systematically repeats these steps at the next map pixel.

\begin{figure}[htb]
\centering
\includegraphics[height=8in, keepaspectratio=true]{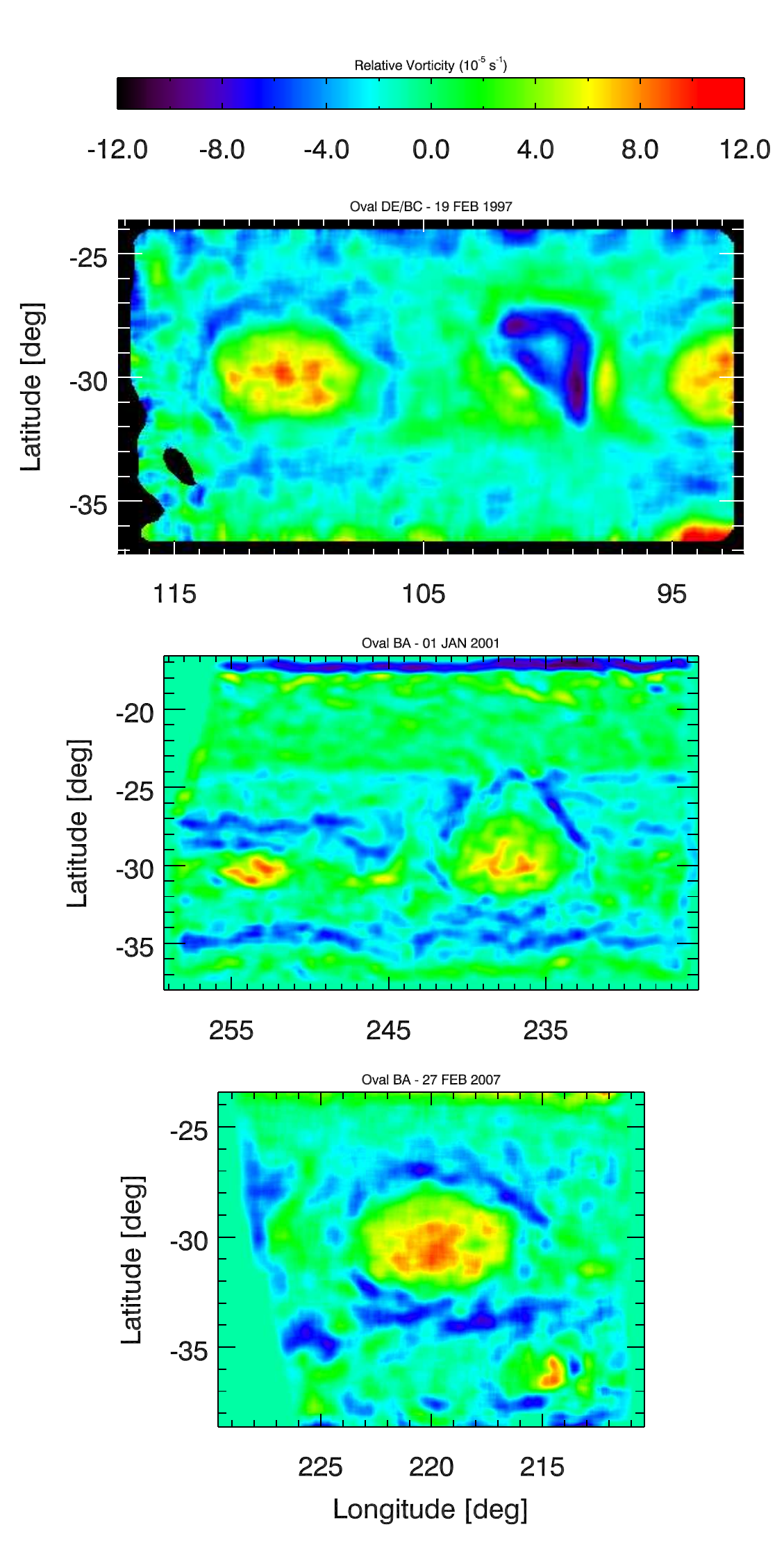}
\caption[Relative Vorticity Maps for Oval DE/BA]{\label{lastfig}
	\label{Figure: relvortmap}
	Map of relative vorticity for Ovals DE/BA from 1997--2007. Positive (negative) values are anticyclonic (cyclonic) areas.
	}
\end{figure}

Figure \ref{Figure: relvortmap} presents a map of relative vorticity between 1997--2007 for Ovals DE and BA. Across all three epochs, the anticyclonic relative vorticity of the Ovals is clearly evident. Examination of the vorticity structure of Oval BA during \emph{Cassini} supplements the argument that the anticyclone was in an atypical state. Both the spatial extent and magnitude of the maximum vorticity anomaly at the central regions of the anticyclones appears to be smaller during the \emph{Cassini} epoch than during the \emph{Galileo} and \emph{New Horizons} observations. This could also be a response to the merger, or a consequence of Oval BA's interaction with the 24$^{\circ}$S zonal jet at its northern boundary during the \emph{Cassini} observations. 

A ring of cyclonic vorticity, similar to the structure found surrounding the Great Red Spot flow collar \citep{Choi07}, can be seen surrounding Ovals DE and BA in Figure \ref{Figure: relvortmap}. In 2007, this cyclonic ring appears to be somewhat stronger and more defined at the southern portion of the cyclone, which is likely associated with the increased velocity within the southern flow collar of Oval BA in combination with the relatively unchanging ambient wind field. The apparent strengthening of the cyclonic ring may also be a signature of stronger downwelling motion in the peripheries of Oval BA. The overall circulation pattern would be thermally indirect \citep{Conrath81}, and suggestive that upwelling and vertical transport is occurring at the core of Oval BA in tandem with the downwelling.   

The cyclone between the two Ovals is clearly apparent in Figure \ref{Figure: relvortmap} as an expansive area of negative relative vorticity. However, the vorticity structure of this cyclone is somewhat peculiar. The strongest areas of vorticity are concentrated along its flow collar whereas its core exhibits weaker negative vorticity. This is opposite in structure compared with the White Ovals and Oval BA, whose relative vorticity weakens with increasing distance away from their centers. Whether or not this structure is typical of Jovian cyclones is unclear. We note that \citet{Hatzes81} measured the flow in a long-lived cyclonic barge in the tropics to reveal a structure similar to what we have measured (a relatively calm center with cyclonic flow at its boundary). The cyclonic feature west of Oval BA during the \emph{Cassini} observations exhibits this same structure in its flow and vorticity, though its relative vorticity appears to be weaker in magnitude along its collar compared with the cyclone imaged in the \emph{Galileo} data set. In any case, additional studies of the detailed flow dynamics of Jovian cyclones are required before determining the prevalence of this structure among all cyclonic Jovian vortices.

\section{Discussion}

The recent reddening of Oval BA has sparked a burst of attention towards the vortex. Multiple works, this one included, are attempting to untangle the truth behind the transformation with the precious number of available data sets useful for relatively precise analysis through automated cloud tracking techniques. Unfortunately, it has become relatively difficult to build a consensus from the collection of recent literature regarding Oval BA's dynamics, as there are apparent contradictory conclusions between several published and ongoing studies. Furthermore, these comparisons between studies are complicated by the varying metrics that authors employ in analyzing the vortex's dynamics, ranging from reporting the measurement of the fastest \emph{individual} wind vector to reporting the highest value of the velocity profile generated from an aggregate of discrete measurements. There have been formal and informal discussions that inadvertently compare these two quantities to each other, when this is not a valid comparison. 

Table \ref{Table: ovalba_comparison} is a compilation of the major dynamical characteristics of Oval BA before and after its color change from an inspection of recent and pending literature. We present comparisons of the wind profiles generated by the series of studies as well as the peak individual wind velocities reported in these works. The quantities for the velocity profile peaks correspond to the maximum value of the zonal or meridional wind profile at each quadrant of Oval BA. The individual measurement peaks are the highest reported wind values made anywhere throughout the vortex; these have been reported previously as simply ``velocities,'' ``peak magnitudes,'' or ``tangential velocity.''

\begin{table}
\begin{center}

\begin{tabular}{lccccl}

\multicolumn{3}{l}{\textbf{Velocity Profile Peaks}} & \multicolumn{3}{}{}\\ 
\hline \hline
Data set & $u$ (N) & $u$ (S) & $v$ (W) & $v$ (E) & Source\\ \hline
\emph{Cassini} (2000a) & -64.2 & 114.8 & -79.4 & 91.4 & \citet{Asay-Davis09}\\
\emph{Cassini} (2000b) & -70 & 110 & -80 & 90 & \citet{Hueso09}\\
\emph{Cassini} (2001) & -72 & 101 & -90 & 98 & \citet{Sussman09}\\
\emph{Cassini} (2001) & -66.8 & 117 & -66.9 & 85.4 & this work\\
\textbf{Average} & \textbf{-68.3} & \textbf{110.7} & \textbf{-79.1} & \textbf{91.2} & \\
\hline
\emph{HST} (2006) & -84.7 & 126.2 & -90.7 & 95.9 & \citet{Asay-Davis09}\\
\emph{HST} (2006) & -90 & 125 & -80 & 80 & \citet{Hueso09}\\
\emph{New Horizons} (2007) & -80 & 115 & -85 & 80 & \citet{Hueso09}\\
\emph{New Horizons} (2007) & -95 & 105 & -97 & 99 & \citet{Sussman09}\\
\emph{New Horizons} (2007) & -85.5 & 138.9 & -85.2 & 78 & this work\\
\textbf{Average} & \textbf{-87.0} & \textbf{122.0} & \textbf{-87.6} & \textbf{86.6} & \\
\hline

\end{tabular}

\vspace{1cm}
(continued next page)

\end{center}
\end{table}

\begin{table}
\begin{center}

\begin{tabular}{lcl}

\multicolumn{3}{l}{\textbf{Individual Measurement Peaks}} \\
\hline \hline
Data set & Velocity & Source \\ \hline
\emph{Cassini} (2000b) & 135 & \citet{Hueso09}\\
\emph{Cassini} (2001) & 132 & \citet{Sussman09}\\
\emph{Cassini} (2001) & 165 $\pm$ 45 & \citet{Simon-Miller06}\\
\emph{Cassini} (2001) & 122.5 & this work\\
\textbf{Average} & \textbf{138.6} & \\
\hline
\emph{HST} (2006) & 180 $\pm$ 70 & \citet{Simon-Miller06}\\
\emph{HST} (2006) & 130 & \citet{Hueso09}\\
\emph{New Horizons} (2007) & 172 $\pm$ 18 & \citet{Cheng08a}\\
\emph{New Horizons} (2007) & 135 & \citet{Hueso09}\\
\emph{New Horizons} (2007) & 127 & \citet{Sussman09}\\
\emph{New Horizons} (2007) & 150.6 & this work\\
\textbf{Average} & \textbf{149.1} & \\
\hline

\end{tabular}

\caption[Comparison of values from recent works regarding Oval BA's dynamics]{
	\label{Table: ovalba_comparison}\label{lasttable}A comparison of values from recent works examining the dynamics of Oval BA before and after its reddening. All velocities are in m s$^{-1}$. We also average the pre-reddened and reddened measurements for comparison. Note that similar, but non-identical, methods are used in averaging the data when constructing profiles for each study.}
\end{center}
\end{table}

From Table \ref{Table: ovalba_comparison}, we see that several works, including our own, consistently show a $\sim$20 m s$^{-1}$ difference before and after the color change in the peak value of the zonal wind profile in the northern half of the vortex; thus, one consensus established by the armada of works regarding Oval BA's dynamics is that the dynamics of the vortex definitely transformed in the northern portion of the vortex in the $\sim$5 years between observations. This is most likely a consequence of the decreasing latitudinal extent of the vortex and the subsequent cessation of its interaction with the 24$^{\circ}$S zonal jet. The consensus for the remaining quadrants is somewhat more difficult to demonstrate. It appears that the meridional wind along Oval BA's eastern portion has experienced a slight decrease in speed between observational epochs. Furthermore, there are also clues suggesting that both Oval BA's meridional wind on its western edge, and its zonal wind on its southern flank, have slightly increased in magnitude. Unfortunately, the evidence for these changes is not as conclusive as the evidence for the change seen at the vortex's northern quadrant before and after the color change (i.e. \emph{Cassini} to \emph{New Horizons} in the peak value, as individual studies suggest either no changes or a changes opposite to that suggested by the statistical trend. In addition, the magnitude of the differences in the velocity profile peaks between the two observational eras may not appear to be significantly more than the amount of measurement uncertainty present. However, it is important to remember that all of these profiles are aggregates of tens to hundreds of individual measurements, each with its own amount of uncertainty, at every point along the latitudinal or longitudinal profile. The ensemble of discrete measurements creating the profile itself would reduce the uncertainties in the values of the velocity profile themselves, implying that these differences in the profiles should not be automatically dismissed as statistically insignificant. Overall, our comparative study examining velocity profiles suggests that Oval BA experienced a modest and spatially non-uniform change in its dynamics sometime between 2001 and 2006, which may or may not have been associated with the coloration event.

Another way of comparing the dynamics of the vortex before and after its color change involves examining the peak individual wind vector measurements. This method has been used historically to characterize vortices and measurements of their winds. Unfortunately, it is typically not as statistically robust as the comparison of velocity profiles since a single measurement is susceptible to the random errors inherent within the measurement technique. Despite this shortcoming, we proceed with exploring this metric here. The broad picture emerging when examining peak individual measurements reported in the published and ongoing studies to date is that it appears unlikely that Oval BA experienced a \emph{significant} acceleration of the flow between the two observational eras before and after the coloration event that resulted in very high wind velocities ($>$ 170 m s$^{-1}$). When solely examining the reports from \citet{Hueso09} and \citet{Sussman09}, both report peak individual measurements that remain fairly consistent between observation eras. Both \citet{Simon-Miller06} and \citet{Cheng08a} report inherently higher values for their velocity measurements, but their reported velocities remain relatively steady before and after the vortex's reddening. (However, these higher values would represent a significant strengthening in Oval BA's flow compared to typical values of the White Ovals' velocity during \emph{Voyager} and \emph{Galileo}.) In contrast with the other studies, our results suggest that Oval BA experienced a moderate increase in its wind speed between observation eras that is mostly confined to the southern portion of its flow collar, as seen in Figure \ref{Figure: velmap}. However, our measurements indicate that the typical flow of the vortex is somewhat slower than the high velocities previously reported by \citet{Simon-Miller06} or \citet{Cheng08a}. 

Thus, the comparison of individual measurement peaks may not be as useful of an approach for characterizing the overall dynamical state of a vortex and what changes (if any) the vortex may have experienced. Because these are single, individual measurements selected from a collection of hundreds or thousands of wind vectors, the direct comparison of them across studies may be inadvertently deceptive and lead to somewhat inaccurate conclusions. We must be cautious in how these measurements are compared with each other. For example, the overall conclusion drawn from a comparison that only highlights the peak individual wind measurements in our current study (the velocity changed moderately) is contradictory with the conclusions drawn from either a sole comparison of individual measurement peaks from other studies (the velocity remained consistent) and with the conclusion drawn from inspecting the velocity profile peaks across various studies (the velocity experienced a modest change that was non-uniform throughout the vortex). It would also be somewhat precarious to conclude that the vortex's dynamics have remained constant simply by reporting the apparent stability of the individual measurement peaks between observations. Comparison of individual measurement peaks may be an oversimplification of the vortex's dynamical meteorology because it omits a large fraction of the results, especially given the current generation of automated techniques that yield a copious amount of wind vectors. One could imagine a scenario where two velocity distributions or profiles have different mean values and standard deviations, but both distributions coincidentally having the same maximum velocity value in its data set. Thus, when possible, we advocate velocity profiles as a primary benchmark in comparative studies of atmospheric phenomena from different observations. The ideal scenario, however, is the comparison of full two-dimensional wind vector maps that reveal the structure of the circulation contained within these atmospheric features.

Apart from its dynamical evolution, Oval BA experienced a remarkable transformation in its visible appearance before and after its color change apart from its hue. The Oval's overall shape as of \emph{New Horizons} bears a strong resemblance to the typical White Oval form from the \emph{Voyager} and \emph{Galileo} eras. The one major difference is the presence of chromophores within the vortex producing a general reddish color overall to the vortex. Observations with HST using various spectral filters indicated that the same coloring agent is likely responsible for the color of both the GRS and Oval BA \citep{Simon-Miller06}. Further spectral analysis concluded that Oval BA extends high into the upper troposphere, also like the GRS. Note that the zonal belts on Jupiter are also colored dark red or brown, but this appearance is likely attributed to a thick, blue-wavelength absorbing or scattering haze overlying clouds at a deeper altitude than clouds at the zones, which are colored white \citep{Simon-Miller01}. Because cyclonic features on Jupiter also appear to be red, it is likely that an unknown coloring agent originating at depth is responsible for the coloring of the vortices \citep{Simon-Miller06}. Furthermore, \citet{Sanchez-Lavega08} observed a towering plume in the northern tropical latitudes of Jupiter during a planetary-scale outburst in March 2007. This plume left in its wake a disturbance that contained red aerosols, which promotes the idea that the coloring agent responsible for the hues in Jupiter's visible appearance originate in the abyssal atmosphere. \citet{Perez-Hoyos09} have recently performed studies regarding the spectral properties of the chromophores within the vortex using standard cloud models of Jovian anticyclones and multispectral data from the Hubble Space Telescope. Their study finds that the cause of the color change in Oval BA is most likely associated with a change in the composition and single scattering albedo of the particles rather than a change in the size distribution of particles within the vortex. In addition, their study concludes that the vertical transport of either inherently red particles or a compound that reacts with incident solar flux and darkens when reaching the upper troposphere is most likely responsible for the coloration of Oval BA. However, this vertical transport would not grossly affect many of the physical and dynamical properties of the vortex, and is likely to be fairly subdued in comparison with typical horizontal mixing and transport rates simultaneously occurring in the atmosphere.

Though evidence for vigorous and significant change in Oval BA's dynamics before and after its color change is beginning to diminish, we believe that there is evidence for a more modest and subdued evolution. Furthermore, this moderate change in the dynamics may have led to the vertical transport of colored particles or gases that condensed or reacted into something which subsequently caused the coloration. When considering what role dynamics had in the reddening of Oval BA, it naturally leads to questions regarding the energy source for the velocity increase. In the shallow-water approximation, potential vorticity ($q$) is defined as 

\begin{equation}
q = \frac{\zeta + f}{h}
\end{equation}

where $h$ is the fluid thickness. If potential vorticity is conserved, an increase in magnitude of $\zeta$ through a spinning up of the vortex would result in an increase in $h$ (i.e. a thicker fluid layer), implying vertical transport both upward and downward within the fluid column through column stretching. Isolated vortices tend to lose energy via Rossby wave radiation \citep{Rhines75}, an interaction that can eventually destroy coherent vortices. \citet{Theiss06} demonstrated that at specific latitudes on Jupiter, this effect can be suppressed, allowing maintenance of long-lived vortices; the White Ovals and Oval BA exist at one of these special latitudes. Thus, the prevention of Rossby wave radiation by Jovian vortices likely plays a role in maintaining the vortices against decay. However, additional energy sources are also thought to contribute towards vortex longevity, and perhaps could have played a role in spinning up Oval BA. Absorption of smaller vortices by larger ones has been postulated \citep{Vasavada05} as one method for sustaining the larger vortex's flow (though this would require the smaller vortices possessing greater potential vorticity than the larger vortex). This method seems unlikely as there are no indications of the White Ovals or Oval BA ingesting smaller vortices in the observational record, though it is commonly observed for the GRS. However, because amateur telescopes provided most observations of Jupiter between 2001--2007, it is plausible that these observations did not have the necessary resolution to resolve the small to medium-scale vortices that could have been absorbed by Oval BA. Another method of energy transfer could have originated from latent heat release via moist convection \citep{Barcilon70, Gierasch76, Ingersoll00}. Moist convection somewhat overlaps with vortex merger, as convection could be responsible for creating the compact vortices that potentially merge with larger ones. Independent from vortex merger, however, is the generation of eddies from moist convection that possibly sustain and accelerate the flow in its ambient surroundings without the creation of a coherent vortex. The observations of bright, white cloud patches in cyclonic regions near Oval BA, possibly indicating moist convection and thunderstorms, lends some credence to this idea. Alternatively, latent heat release could perhaps have occurred within large scale ascent across the vortex; this can allow energy release even in the absence of thunderstorm generation \citep{Showman05}. Perhaps the energy released by such latent heating slowly accelerated the flow of Oval BA, gradually causing spin-up of fluid columns, advection and vertical transport of chemical species from deep layers, and eventual coloration. Future numerical modeling studies will assess these energy transport modes and evaluate their applicability to Jovian atmospheric conditions.   

The apparent similarities between the Oval BA and the Great Red Spot is intriguing when we consider that some winds on Oval BA now flow at speeds rivaling those historically observed on the Great Red Spot. It is tempting to speculate that BA is now a more compact version of the Great Red Spot, especially when considering the resemblances in their color and the turbulent wake regions to the vortices' northwest. We hope that continued observation and data analysis of Oval BA and the Great Red Spot coupled with radiative transfer and dynamical modeling of these systems will prove fruitful in further understanding the similarities and differences between both vortices. 

\acknowledgements
We thank Dr. Xylar Asay-Davis and Dr. Agustin S\'anchez-Lavega for their constructive reviews which improved this manuscript. We thank Dr. Amy Simon-Miller and Michael Sussman for thoughtful discussions that improved this work. We thank Nathaniel Nerode for his work in compiling the raw \emph{Galileo} images and producing mosaics for the PDS Atmospheres node. This research was supported by NASA Planetary Atmospheres grants to APS and a NASA Earth and Space Science Fellowship, \#NNX08AW01H. Additional support was provided by the University of Arizona TRIF Imaging Fellowship.

\label{lastpage}

\bibliographystyle{elsart-harv-choi} 
\bibliography{ovalba}

\end{document}